# The Power Spectrum of IRAS Galaxies

Helen Tadros and George Efstathiou,
*Department of Physics, University of Oxford, Keble Road, Oxford, OX1 3RH, UK.*

**ABSTRACT**
We estimate the three-dimensional power spectrum of IRAS galaxies from the QDOT and 1.2Jy redshift surveys. We use identical estimators for both surveys and show how the results depend on the weights assigned to the galaxies. The power spectrum for the QDOT survey is steeper and has a higher amplitude at wavenumbers $k \sim 0.05\ h\text{Mpc}^{-1}$ (where $h$ is Hubble's constant in units of $100\text{km s}^{-1}\text{ Mpc}^{-1}$) than the power spectrum derived from the 1.2Jy sample. However, the QDOT power spectrum is sensitive to a small number of galaxies in the Hercules supercluster, in agreement with a recent analysis of galaxy counts in cells in these surveys. We argue that the QDOT results are an upward fluctuation. We combine the two surveys to derive our best estimate of the power spectrum of IRAS galaxies. This is shallower and has a lower amplitude on scales $\lesssim 0.1 h\text{Mpc}^{-1}$ than the power spectrum derived by Feldman *et al.* (1994) from the QDOT survey alone. The power spectrum of the combined surveys is well described by the linear theory power spectrum of a scale-invariant cold dark matter model with $\Omega h = 0.2$.

**Key words:** cosmology: large-scale structure of Universe

## 1 INTRODUCTION

The power spectrum of galaxy clustering provides one of the strongest constraints on theories of the formation of structure in the Universe (see Efstathiou 1995a and Strauss and Willick 1995 for recent reviews). The three dimensional power spectrum of optically selected galaxies has been estimated by a number of authors (see *e.g.* Vogeley *et al.* 1992, Park *et al.* 1994, Baugh and Efstathiou 1993, 1994). The power spectrum of IRAS galaxies has been estimated by Fisher *et al.* (1993) from the 1.2Jy redshift survey of Fisher *et al.* (1995) and by Feldman, Kaiser and Peacock (1994, hereafter FKP) from the QDOT survey of Lawrence *et al.* (1995). These estimates of the IRAS power spectrum differ in shape and amplitude (see *e.g.* Figure 5.7 of Efstathiou 1995a). However, different techniques and weighting schemes were used by these authors and so it is unclear whether the differences in the power spectra are attributable to the estimators or to the galaxy catalogues. In this *Letter* we address this problem by applying the techniques of FKP to the 1.2Jy and QDOT surveys, using identical weighting schemes and sky coverage.

The results presented here are closely related to a statistical analysis of galaxy counts-in-cells in the 1.2Jy and QDOT surveys described by Efstathiou 1995b, hereafter E95). By comparing the counts on a cell-by-cell basis, E95 found that the two surveys were compatible with the hypothesis that they sample the same underlying density field.

The variances in the cell counts of the QDOT survey on scales $\ell \sim 30$–$40\ h^{-1}\text{Mpc}$ are, however, systematically higher than those derived from the 1.2Jy survey. Most of these differences are caused by a small number of galaxies in the QDOT survey that lie in the region of the Hercules supercluster. E95 concluded that the QDOT variances were biased high because Hercules is over-represented in the QDOT survey.

Here we show that similar conclusions apply to the power spectra of IRAS galaxies. The techniques and surveys are described in Section 2. We investigate the sensitivity of the power spectrum estimates to the weights assigned to the galaxies and we show that the QDOT power spectrum is sensitive to galaxies in the Hercules supercluster and, consequently, to the weighting scheme. In contrast, the power spectra derived from the 1.2Jy survey are insensitive to galaxies in Hercules and the weighting scheme. We combine the two surveys to produce our best estimate of the power spectrum of IRAS galaxies. Our conclusions are presented in Section 3 together with a brief comparison of our results with cold dark matter (CDM) models.

## 2 ESTIMATION OF THE POWER SPECTRUM

Both the QDOT survey and the 1.2Jy survey are based on version 2 of the IRAS Point Source Catalogue (PSC). The QDOT survey (Lawrence *et al.* 1995) is a 1 in 6 sparse-sampled redshift survey of galaxies with 60$\mu$m fluxes greater than 0.6Jy, while the 1.2Jy survey is fully sampled at a brighter flux limit of 1.2Jy (Fisher et al. 1995). The QDOT catalogue has been corrected for the 211 erroneous redshifts in the original version. Both surveys exclude regions of the sky (defined by a sky mask) at low galactic latitudes and regions of high Galactic emission at 60$\mu$m (see Strauss et al. 1990, Rowan-Robinson et al. 1991). For the power spectrum analysis presented here we have used a concatenated QDOT and 1.2Jy sky mask and used galaxies within 450 $h^{-1}$Mpc that lie above a galactic latitude $|b| = 10°$. This leaves 2024 galaxies in the QDOT sample and 4384 galaxies in the 1.2Jy sample.

We have followed the methods described in FKP to compute the power spectrum of a flux limited redshift survey. To simplify the discussion we use the same notation as FKP and refer to their paper for details of the method. We compute a weighted density field

$$F(\mathbf{r}) = \frac{w(r)[n_g(\mathbf{r}) - \alpha n_s(\mathbf{r})]}{\left[\int \overline{n}^2(\mathbf{r}) w^2(r) d^3r\right]^{\frac{1}{2}}}, \quad (1)$$

where the subscript $g$ denotes the galaxy density in the real catalogue and $s$ denotes the density field for a random catalogue with same angular and radial selection functions as the galaxy survey, computed from equations (4) and (5) below. In equation (1), $\overline{n}(\mathbf{r})$ is the expected mean density of galaxies in a catalogue with the same angular and radial selection functions as the data. The function $\overline{n}(\mathbf{r})$ is therefore the mean galaxy density $\overline{n}(r)$ multiplied by the angular mask. The factor $\alpha$ is the ratio of the space densities in the real catalogue to that in the random catalogue. In the analysis presented here we use several hundred times as many points in the random catalogue as there are galaxies in the real catalogue, and compute $\alpha$ from the ratio of the sums

$$\sum_i \frac{1}{(1 + 4\pi \overline{n}(r_i) J_3)}, \quad (2)$$

where we have set $4\pi J_3 = 10000(h^{-1}\mathrm{Mpc})^3$ and the sums run over all galaxies and random points. Equation (2) provides a minimum variance estimate of the mean galaxy density and the specific choice for $J_3$ is based on the power spectra of CDM models, though the results presented below are insensitive to this value (see Efstathiou 1995a, section 5.3 for details).

The weight function, $w(r)$ that minimises the variance of the power spectrum $P(k)$, under the assumption that the fluctuations are Gaussian, is given by

$$w(r) = \frac{1}{1 + \overline{n}(r) P(k)}, \quad (3)$$

(FKP) and so depends on the value of $P(k)$ at each wavenumber $k$. As in FKP we have decided to set $P(k)$ in equation (3) equal to a constant value and to show how the estimates of the power spectrum change for four values of $P(k) = 2000, 4000, 8000$ and $16000$ $(h^{-1}\mathrm{Mpc})^3$, which span the range of interest. Changing the value of $P(k)$ used in equation (3) changes the effective depth of the catalogue. A larger value of $P(k)$ results in a larger effective depth. The power spectrum is derived by Fourier transforming equation (1), removing shot noise (equation 2.1.9 of FKP) and averaging the power-spectrum estimates over shells in $k$-space of volume $V_k$. The power spectrum estimates will be correlated over a range in k space give approximately by $\delta k \sim D^{-1}$ where D is the characteristic depth of the survey.

We estimate the mean galaxy density from the luminosity function $\phi(L) dL$ for IRAS galaxies

$$\overline{n}(r) = \int_{L_{min}(r)}^{\infty} \phi(L) \, dL, \quad (4)$$

where $L_{min}(r)$ is the luminosity of a galaxy at distance $r$ with flux equal to the flux limit of the survey. Galaxy distances are calculated from redshifts using the relativistic formula and assuming the Hubble constant is 100 km s$^{-1}$ Mpc$^{-1}$ and $\Omega$ equals 1. We use the parametric form of the luminosity function given by Saunders et al. al (1990),

$$\phi(L) = C \left(\frac{L}{L_\star}\right)^{(1-\alpha)} exp\left[-\frac{1}{2\sigma^2} log_{10}^2 \left(1 + \frac{L}{L_\star}\right)\right], \quad (5)$$

with parameters

$C = \frac{2.6}{log(10)} \times 10^{-2} \ h^3\mathrm{Mpc}^{-3}$
$\alpha = 1.09$
$\sigma = 0.724$
$L_\star = 10^{8.47} h^{-2} L_\odot$.

Because of the relatively small median depth of the samples, evolutionary effects in $\phi(L)$ are unimportant compared to the random errors in $P(k)$ since they affect only the high redshift tail of the $\overline{n}(r)$ distribution. The parameters given above provide a very good fit to the $\overline{n}(r)$ for both the QDOT and the 1.2Jy surveys.

In Figure 1 we show the power spectrum of the QDOT survey (circles) and the 1.2Jy survey (crosses) for each of the four values of $P(k)$ used in the weighting function. The QDOT spectra shown in Figure 1 are in very good agreement with those published by FKP. Our estimates of the power spectrum for the 1.2Jy survey are also in good agreement with those of Fisher et al. (1993), but a point-by-point comparison is not straightforward because they use a different weighting scheme and an estimator which returns the power spectrum convolved with a cylindrical window function. Figure 1 illustrates the sensitivity of the estimates to the weighting scheme. As we increase the value of $P(k)$ in equation 3, we assign progressively higher weight to more distant galaxies. Since the weights are designed to give a minimum variance estimate of $P(k)$, we would not expect the results to be sensitive to small changes in the weighting, and indeed the results for the 1.2Jy survey shown in Figure 1 are insensitive to changes in the weighting. However, the power spectrum measured from the QDOT survey is much more sensitive to the weights and increases systematically as we give proportionately higher weight to more distant galaxies. The estimates from the two surveys are consistent with each other when we weight with $P(k) = 2000(h^{-1}\mathrm{Mpc})^3$, as noted by FKP. However, when we weight with $P(k) \geq 8000(h^{-1}\mathrm{Mpc})^3$, the QDOT power spectrum lies about a factor of 2 or 3 higher than that of the 1.2Jy survey over the wavenumber range $0.08 \gtrsim k \gtrsim 0.03 h\mathrm{Mpc}^{-1}$. From the amplitudes of the power spectra plotted in Figure 1, we would expect that weighting with $P(k) \sim 8000(h^{-1}\mathrm{Mpc})^3$ would be close to optimal over the wavenumber range 0.05–0.1$h\mathrm{Mpc}^{-1}$. It is surprising, therefore, that the power spectra estimated from the QDOT survey are so sensitive to the weights and appear to differ systematically from those of the 1.2Jy survey when we weight with $P(k) \geq 8000(\ h^{-1}\text{Mpc})^3$.

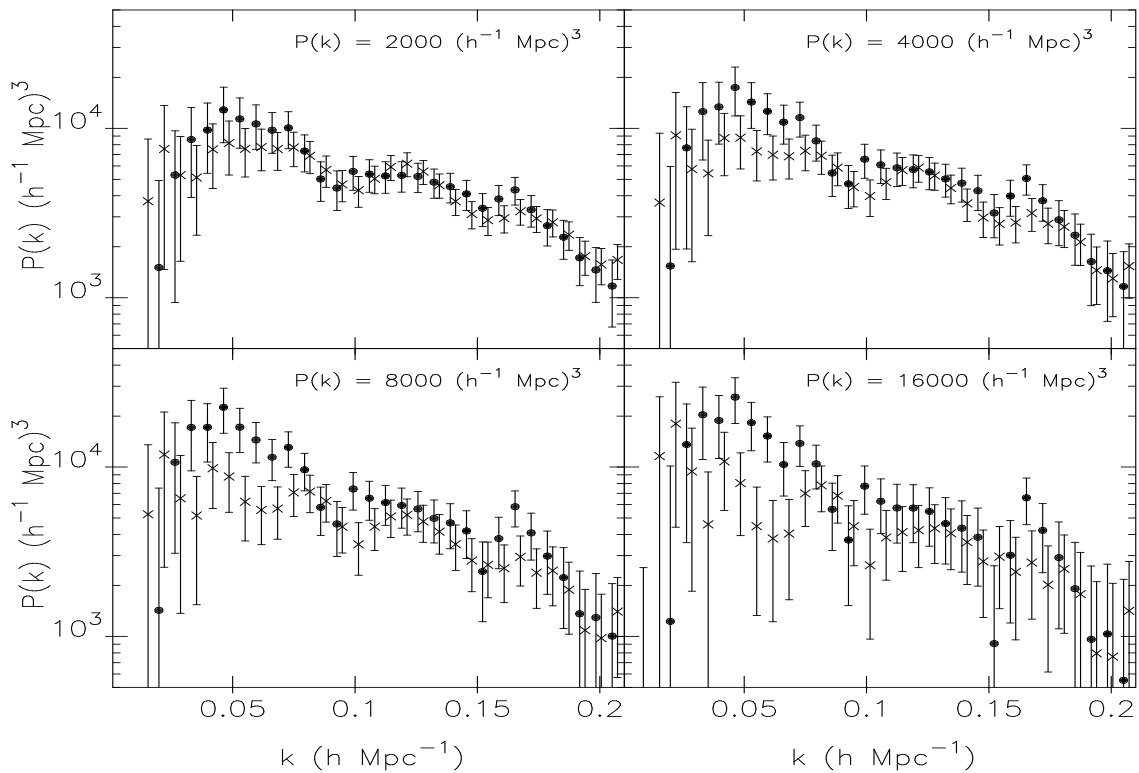

**Figure 1.** Power spectra of IRAS galaxies determined with four weighting schemes, $P(k) = 2000\text{ -}16000\ (h^{-1}\text{Mpc})^3$ in equation 3, as indicated in each panel. The crosses show the power spectra measured from the 1.2Jy survey and the filled circles show the power spectra measured from the QDOT survey. The error bars show one standard deviation, computed from equation 2.4.6 of FKP. For clarity, the IRAS 1.2Jy power spectra have been shifted to the right of the plot by a third of the spacing between points in k-space.

The analysis of the 1.2Jy and QDOT surveys described in E95, shows that the cell statistics of the QDOT survey are extremely sensitive to a small number of galaxies in the region of the Hercules supercluster ($l = 48°$, $b = 43°$, $v = 11000\text{km s}^{-1}$). A similar conclusion applies to the power spectrum analysis described here. We illustrate this point by recomputing the power spectra after removing galaxies in the Hercules region. The QDOT survey was divided into 54 regions of the sky called q-sectors (see Lawrence *et al.* 1995). The Hercules supercluster is located largely in q-sector 7 which covers the area $40° \leq b \leq 60°$, $0° \leq l \leq 60°$. We exclude Hercules by removing all galaxies that lie in q-sector 7; this removes 59 galaxies from the QDOT survey and 74 galaxies from the 1.2Jy survey. Figure 2 shows the power spectra after this is done. The power spectra for the 1.2Jy survey are almost identical to those plotted in Figure 1, showing that Hercules has little effect on this survey. In contrast, the results for the QDOT survey differ significantly from those plotted in Figure 1. With the Hercules region removed, the QDOT power spectra are in excellent agreement with those of the 1.2Jy survey and are insensitive to the weighting scheme. There is little doubt, therefore, that the differences between the 1.2Jy and QDOT power spectra apparent in Figure 1, and the sensitivity of the QDOT power spectra to the weighting scheme, are caused by a small number of QDOT galaxies in the Hercules region. This is consistent with the results of the cell count analysis described in E95. Galaxies in the Hercules region are over-represented in the QDOT survey (see E95) and so bias the power spectrum to high values. As the factor $P(k)$ is increased in the weighting function, the Hercules region is given higher weight and so the discrepancy with the 1.2Jy power spectrum becomes more pronounced.

Although the analysis in E95 shows that the Hercules region is over-represented in the QDOT survey, most likely because of a statistical fluctuation in the random numbers used to construct the survey, the 1.2Jy and QDOT surveys were found to be consistent with each other within Poisson statistics. To obtain an estimate of $P(k)$ from the two surveys it therefore seems reasonable simply to add them together without excluding galaxies from any particular region. In practice the 1.2Jy survey carries so much weight in the analysis of the combined surveys that it makes little difference whether or not we exclude Hercules from QDOT. Figure 3 shows the power spectrum derived for the combined surveys. In this Figure, we remove QDOT galaxies with $60\mu$m fluxes above 1.2Jy, since these are already included in the 1.2Jy survey. We used the concatenated QDOT-1.2Jy mask, as in the estimates shown in Figure 1, and a selection function for the combined surveys computed from equations (4) and (5). We have set $P(k) = 8000(h^{-1}\text{Mpc})^3$ in the weighting function of equation (3); varying $P(k)$ in the weight function over the range $2000\text{ -}16000\ h^3\text{Mpc}^{-3}$ shifts the points by less than the $1\sigma$ error bars. The power spectrum plotted in Figure 3 is about a factor of two lower in am-plitude over the wavenumber range $0.03 \lesssim k \lesssim 0.08 h\mathrm{Mpc}^{-1}$ than the QDOT power spectrum plotted in Figure 7 of FKP. In Table 1 we give the values of the power spectra and the one sigma errors for the combined survey, weighted with $P(k) = 4000, 8000,$ and $16000 (h^{-1}\mathrm{Mpc})^3$.

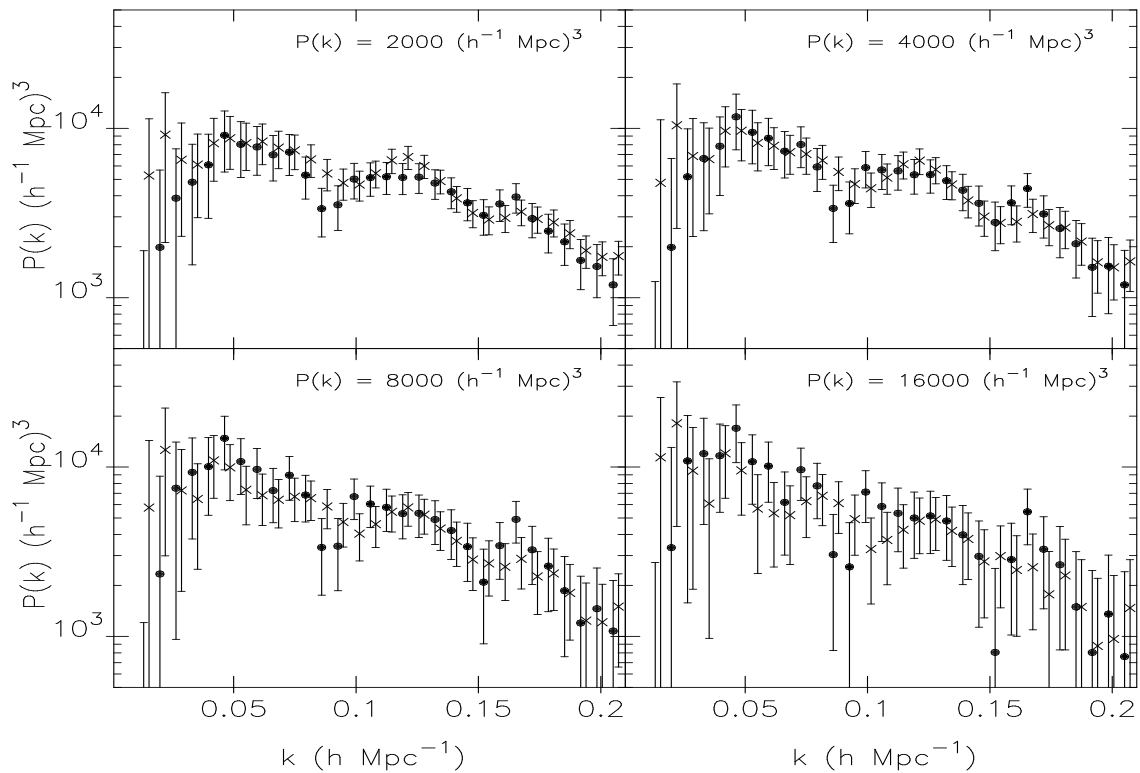

**Figure 2.** As Figure 1 except that here we have removed 59 galaxies from the QDOT survey and 74 galaxies from the 1.2Jy survey that lie in the region of the Hercules supercluster.

## 3 CONCLUSIONS

Our main goal in this Letter has been to compare the power spectra of the 1.2Jy and QDOT surveys using the same estimators and weighting schemes. Our results show that the QDOT power spectrum is sensitive to a small number of galaxies in the Hercules supercluster. The QDOT power spectrum lies systematically higher than the power spectrum of the 1.2Jy survey over the wavenumber range $0.03 \lesssim k \lesssim 0.08 h\mathrm{Mpc}^{-1}$. The power spectrum for the combined surveys plotted in Figure 3 is insensitive to galaxies in Hercules and to the weighting scheme and lies well within the $1\sigma$ errors of the power spectra for the 1.2Jy survey plotted in Figure 1. These results are in accord with the counts-in-cells analysis of the 1.2Jy and QDOT surveys discussed in E95.

The two lines in Figure 3 show the linear theory power spectra of two scale invariant CDM models convolved with the window function of the combined surveys[*]. These curves are computed from equation 7 of Efstathiou *et al.* 1992, for two values of the parameter $\Gamma = \Omega h$, and are normalised so that the *rms* fluctuation in spheres of radius $8 h^{-1}\mathrm{Mpc}$, $\sigma_8$, is equal to unity. We do not attempt a detailed comparison with theoretical models here since this requires an analysis of the distortions to the power spectrum caused by peculiar motions (see Kaiser 1987) and of the relative bias between the clustering of IRAS galaxies and the mass distribution. We note, however, that the general shape of the power spectrum plotted in Figure 3 resembles that of a low density CDM model with $\Gamma = 0.2$ and $\sigma_8 \approx 0.84$ and that there is some evidence for excess power at wavenumbers $k \sim 0.04 h\mathrm{Mpc}^{-1}$ compared to the standard CDM model with $\Gamma = 0.5$ normalised to $\sigma_8 \approx 1$. These results are in qualitative agreement with the conclusions of FKP, and with those deduced from the power spectra of optically selected redshift surveys (*e.g.* Park *et al.* 1994). A comparison of the power spectrum of IRAS galaxies, presented here and optical galaxies is discussed in Efstathiou *et al.* 1995. The authors find the relative bias between optical and IRAS galaxies to be fairly small, $P_{opt}(k)/P_{iras}(k) \approx 1.4$ over the wavenumber range $0.03 \lesssim k \lesssim 0.2 h\mathrm{Mpc}^{-1}$

**Acknowledgments:** Helen Tadros acknowledges the receipt of a PPARC research studentship and George Efstathiou thanks PPARC for the award of a Senior Fellowship. We thank the referee, Michael Strauss, for useful comments.

---

[*] The convolution reduces the amplitude of the theoretical power spectra by about 15% at wavenumbers $\lesssim 0.1 h\mathrm{Mpc}^{-1}$.

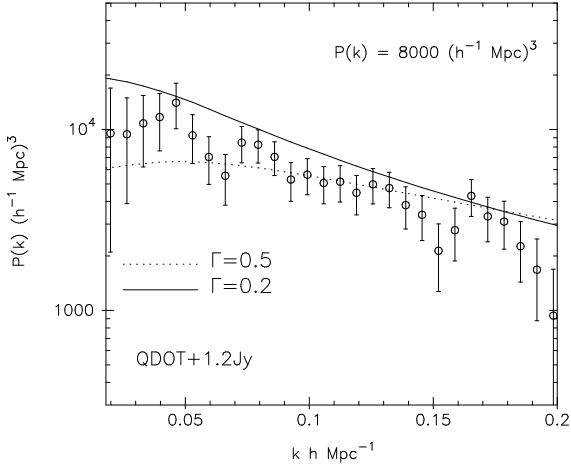

**Figure 3.** The power spectrum of the combined QDOT and 1.2Jy surveys. The error bars show one standard deviation and we adopted $P(k) = 8000(h^{-1}\mathrm{Mpc})^3$ in the weight function (equation 3). The two lines show the linear theory power spectra for two scale-invariant CDM models with $\Gamma = \Omega h = 0.2$ and $0.5$. The linear theory curves have been normalised so that the *rms* fluctuation in spheres of radius $8\ h^{-1}\mathrm{Mpc}$ is equal to unity.

| Weight | 4000 | 8000 | 16000 |
|---|---|---|---|
| $k$ | $P(k)$ | $P(k)$ | $P(k)$ |
| 0.0198 | 5346 ± 4689 | 9519 ± 7420 | 15990 ± 11552 |
| 0.0265 | 6216 ± 3858 | 9418 ± 5528 | 13890 ± 7992 |
| 0.0331 | 8744 ± 3700 | 10817 ± 4607 | 13535 ± 6094 |
| 0.0397 | 10474 ± 3540 | 11708 ± 4094 | 12049 ± 4949 |
| 0.0463 | 12748 ± 3541 | 14062 ± 3968 | 15103 ± 4755 |
| 0.0529 | 9611 ± 2581 | 9278 ± 2794 | 8709 ± 3397 |
| 0.0595 | 7713 ± 1918 | 7047 ± 2078 | 5940 ± 2549 |
| 0.0661 | 6849 ± 1623 | 5546 ± 1730 | 3005 ± 2076 |
| 0.0728 | 8639 ± 1735 | 8464 ± 1922 | 8096 ± 2389 |
| 0.0794 | 8133 ± 1515 | 8243 ± 1714 | 8066 ± 2145 |
| 0.0860 | 6881 ± 1266 | 7057 ± 1490 | 7467 ± 1970 |
| 0.0926 | 5536 ± 1068 | 5285 ± 1283 | 5685 ± 1787 |
| 0.0921 | 5693 ± 1042 | 5628 ± 1270 | 5828 ± 1750 |
| 0.1058 | 5254 ± 953 | 5059 ± 1174 | 4677 ± 1609 |
| 0.1124 | 5366 ± 942 | 5143 ± 1170 | 4618 ± 1609 |
| 0.1190 | 5074 ± 894 | 4473 ± 1099 | 3897 ± 1540 |
| 0.1257 | 5208 ± 876 | 4980 ± 1102 | 5223 ± 1579 |
| 0.1323 | 4846 ± 816 | 4746 ± 1055 | 5547 ± 1564 |
| 0.1389 | 3922 ± 744 | 3820 ± 1000 | 4304 ± 1508 |
| 0.1455 | 3309 ± 674 | 3375 ± 941 | 3702 ± 1432 |
| 0.1521 | 2490 ± 615 | 2139 ± 867 | 1893 ± 1337 |
| 0.1587 | 2849 ± 632 | 2776 ± 898 | 2996 ± 1391 |
| 0.1653 | 3711 ± 687 | 4295 ± 989 | 4999 ± 1502 |
| 0.1720 | 3016 ± 627 | 3308 ± 913 | 3517 ± 1400 |
| 0.1786 | 2833 ± 623 | 3098 ± 917 | 3500 ± 1430 |
| 0.1852 | 2303 ± 563 | 2264 ± 829 | 2296 ± 1302 |
| 0.1918 | 1734 ± 528 | 1679 ± 802 | 2068 ± 1304 |
| 0.1984 | 1132 ± 488 | 935 ± 755 | 1236 ± 1251 |
| 0.2050 | 1229 ± 497 | 1095 ± 770 | 1441 ± 1276 |

**Table 1.** Numerical values for the power spectrum of the combined QDOT and 1.2Jy survey for three different values of $P(k)$ in the weighting scheme. In the first line, weight refers to the value of $P(k)$ used in the weighting scheme (Equation 3). Units of k are $h\mathrm{Mpc}^{-1}$ and units of $P(k)$ are $\left(h^{-1}\mathrm{Mpc}\right)^3$. The points in the power spectrum estimate are correlated, (see equation 2.5.2 of FKP) over a range $\delta k \approx 0.011$ for the weightings shown in this table, i.e about one and a half times the spacing between points.